\documentclass[cernpreprint, atlasdraft=false, texlive=2025, UKenglish,REPORT=false, texmf, orcidlogo]{atlasdoc}
\usepackage{atlaspackage}
\usepackage{atlasbiblatex}

\usepackage{atlascontribute}

\usepackage{ifthen}
\usepackage{mathrsfs}
\usepackage{rotating}
\usepackage{amsmath}
\usepackage{multirow}
\usepackage{dcolumn}

\usepackage{siunitx}
\usepackage{atlasphysics}
\usepackage{tikz}
\usetikzlibrary{calc,math}
\usepackage{xfp}

\usepackage{multicol}

\graphicspath{{logos/}{figures/}{figures/internal/}}

\newboolean{InternalLinksInPublic}
\setboolean{InternalLinksInPublic}{false}   %

\newcommand{\tab}{Table~}
\newcommand{\fig}{Figure~}
\newcommand{\sect}{Section~}
\newcommand{\eq}{Eq.~}

\newcommand{\CoMEnergy}{13~\TeV}

\newcommand{\CoMEnergySeven}{7~\TeV}
\newcommand{\ilumi}{$ 140~\mathrm{fb^{-1}} $}

\newcommand{\Jpsitomm}{\ensuremath{ \Jpsi \ra \mumu }\xspace} 			%

\newcommand{\KstarKPi}{\ensuremath{K^{*0}\to K^+ \pi^-}\xspace} 			%

\newcommand{\chisq}{$\mathrm \chi^{2}$/ndof }
\newcommand{\cutChiBd}{\chisq $< 3$}
\newcommand{\cutChiJpsi}{\chisq $< 10$}

\newcommand{\ntotalBdtext}{\mbox{10\,559\,554}}			%
\newcommand{\cutMassKstar}{$ 846 \; \textrm{MeV} < m(K^+ \pi^-) < 946 \; \textrm{MeV} $}

\newcommand{\cutMassBdRange}{5.00--5.65~\GeV}

\newcommand{\Jpsipa}{\Jpsi \ }

\newcommand{\Bp}{\ensuremath{B^{+} }}
\newcommand{\Bo}{\ensuremath{B^{0} }} %

\newcommand{\Bso}{\ensuremath{B^{0}_{s} }}
\newcommand{\Kns}{\ensuremath{K^{*0} }} %
\newcommand{\Knsbar}{\ensuremath{\bar{K}^{*0} }} %

\newcommand{\JpsiKnst}{\ensuremath{  J/\psi\Kns } }

\newcommand{\JpsiK}{\ensuremath{ J/\psi K^{+} }\space}

\newcommand{\BtoKns}{\ensuremath{  \Bo \ra  J/\psi  \Kns}\xspace}					%

\newcommand{\BtoKs}{\ensuremath{  \Bo \ra  \Jpsi  \KzeroS}} %

\newcommand{\ppJpsiX}{\ensuremath{ pp \ra \Jpsi X  }}

\newcommand{\BpToJpsiK}{\ensuremath{\Bp \ra \JpsiK}}
\newcommand{\BoToJpsiKnst}{\ensuremath{\Bo \ra \JpsiKnst}}

\providecommand{\tildet}{\ensuremath{t}\xspace}

\providecommand{\tildeti}{\ensuremath{t_i}\xspace}
\providecommand{\sigtildeti}{\ensuremath{\sigma_{t_{i}}}\xspace}

\tikzmath{
\tauvalue = 1.5053;
\tauBdStat = 0.0012;
\tauBdSystA = 0.001085; %
\tauBdSystB = \fpeval{round(0.87*0.0012,6)};
\tauBdSystC = 0.000001;
\tauBdSystD = \fpeval{round(sqrt(0.00092^2+0.0007^2+0.0006^2 + 0.00036^2),6)};
\tauBdSystE = \fpeval{round(0.0012*sqrt(1.00^2+0.42^2+0.66^2),6)};
\tauBdSystF = 0.00002;
\tauBdSystJ = 0.0002;
\tauBdSystK = \fpeval{round(0.34*0.0012,6)};
\tauBdSystG = \fpeval{round(sqrt(0.5*0.0012*0.5*0.0012+0.3*0.0012*0.3*0.0012),6)};
\tauBdSystH = \fpeval{round(0.084*0.0012,6)};
\tauBdSystL = \fpeval{round(1.91*0.0012,6)};
\tauBdSystToTableA = \fpeval{round(\tauBdSystA,5)};
\tauBdSystToTableB = \fpeval{round(\tauBdSystB,5)};
\tauBdSystToTableC = \fpeval{round(\tauBdSystC,5)};
\tauBdSystToTableD = \fpeval{round(\tauBdSystD,5)};
\tauBdSystToTableE = \fpeval{round(\tauBdSystE,5)};
\tauBdSystToTableF = \fpeval{round(\tauBdSystF,5)};
\tauBdSystToTableG = \fpeval{round(\tauBdSystG,5)};
\tauBdSystToTableH = \fpeval{round(\tauBdSystH,5)};
\tauBdSystToTableJ = \fpeval{round(\tauBdSystJ,5)};
\tauBdSystToTableK = \fpeval{round(\tauBdSystK,5)};
\tauBdSystToTableL = \fpeval{round(\tauBdSystL,5)};
\tauBdSyst = \fpeval{round(sqrt(\tauBdSystA^2 + \tauBdSystB^2 + \tauBdSystC^2 + \tauBdSystD^2 + \tauBdSystE^2 + \tauBdSystF^2+\tauBdSystG^2+\tauBdSystH^2+\tauBdSystJ^2+\tauBdSystK^2+\tauBdSystL^2),4)};
\tauDifftoHflav =  \fpeval{round( abs(\tauvalue - 1.517) / sqrt(\tauBdSyst^2 + \tauBdStat^2+ 0.004^2),1) }; %
\tauDifftoLHCbKS =  \fpeval{round(abs(\tauvalue - 1.499) / sqrt(\tauBdSyst^2 + \tauBdStat^2+ 0.013^2+ 0.005^2),1) };
\tauDifftoLHCbKsta =  \fpeval{round( abs(\tauvalue - 1.524) / sqrt(\tauBdSyst^2 + \tauBdStat^2+ 0.006^2+ 0.004^2),1) };
\tauDifftoLHCbKpi =  \fpeval{round( abs(\tauvalue - 1.524) / sqrt(\tauBdSyst^2 + \tauBdStat^2+ 0.011^2+ 0.004^2),1) };
\Aperp = 0.211;
\AperpErr = 0.008;
\Azero = 0.571;
\AzeroErr = 0.007;
\yvalue = 0.0005;
\yStatErr = 0.005;
\Apara = 1-\Aperp-\Azero;
\AparaErr = \fpeval{round(sqrt(\AperpErr*\AperpErr+\AzeroErr*\AzeroErr),3)};
\RHvalue = \Aperp;
\RHErr = \AperpErr;
\RLvalue = \Azero + \Apara;
\RLErr = sqrt( \AzeroErr^2 + \AparaErr^2 );
\Avalue =  \fpeval{round((\RHvalue-\RLvalue)/(\RHvalue+\RLvalue) ,4)}; %
\AstatErr = \fpeval{round(2*sqrt((\RHvalue*\RLErr)^2+(\RLvalue*\RHErr)^2)/((\RHvalue+\RLvalue)^2),4)}; %
\Gammavalue = \fpeval{round((1/\tauvalue)*( 1/(1-\yvalue^2) )*(1+\yvalue^2),4)};
\GammaStatErr = \fpeval{round(\Gammavalue*\tauBdStat/\tauvalue,4)};
\GammaSystErr = \fpeval{round(\Gammavalue*\tauBdSyst/\tauvalue + 0.0001,4)};
\GammaExteErr = \fpeval{round(sqrt(((4 *\yvalue + \Avalue + 6* \yvalue^2 *\Avalue + \yvalue^4 *\Avalue + 4* \yvalue^3 *\Avalue^2)/(\tauvalue*(1 - \yvalue^2)^2 *(1 + \yvalue*\Avalue)^2))^2*\yStatErr^2 + (\yvalue/(\tauvalue* (1 + \yvalue *\Avalue )^2))^2*\AstatErr^2),4)};
\GammaSystErrNanosecond = \fpeval{round(1000*\GammaSystErr,1)};
\GammaDSystToTableA = \fpeval{round(1000*\tauBdSystA*\Gammavalue/\tauvalue,2)};
\GammaDSystToTableB = \fpeval{round(1000*\tauBdSystB*\Gammavalue/\tauvalue,2)};
\GammaDSystToTableC = \fpeval{round(1000*\tauBdSystC*\Gammavalue/\tauvalue,2)};
\GammaDSystToTableD = \fpeval{round(1000*\tauBdSystD*\Gammavalue/\tauvalue,2)};
\GammaDSystToTableE = \fpeval{round(1000*\tauBdSystE*\Gammavalue/\tauvalue,2)};
\GammaDSystToTableF = \fpeval{round(1000*\tauBdSystF*\Gammavalue/\tauvalue,2)};
\GammasValue = 0.6703;
\GammasStat = 0.0014;
\GammasSyst = 0.0018;
\ratioValue = \fpeval{round(\Gammavalue/\GammasValue,4)};
\ratioStatErr = \fpeval{round(\ratioValue*sqrt((\GammasStat/\GammasValue)^2 + (\GammaStatErr/\Gammavalue)^2),4)};
\ratioSystErr = \fpeval{round(\ratioValue*sqrt((\GammasSyst/\GammasValue)^2  + (\GammaSystErr/\Gammavalue)^2),4)};
\ratioExteErr = \fpeval{round(\ratioValue*\GammaExteErr/\Gammavalue,4)};
\RatioDiffHQE =  \fpeval{round( abs( \ratioValue - 1.003 ) / sqrt(\ratioStatErr^2 + \ratioSystErr^2                   +    0.006^2),1) };
\RatioDiffLatice =  \fpeval{round( abs( \ratioValue - 1.00 ) / sqrt(\ratioStatErr^2 + \ratioSystErr^2                   + 0.02^2),1) };
\GammaDifftoHQE =  \fpeval{round( abs(\Gammavalue - 0.63 ) / sqrt(\GammaSystErr^2 + \GammaStatErr^2                  + 0.11^2),1) };
}

\newcommand{\BdTauFitVal}{ \ensuremath{ \tauvalue}} %
\newcommand{\BdTauStatUncrt}{ \ensuremath{ \tauBdStat}}
\newcommand{\BdTauSystUncrt}{ \ensuremath{ \tauBdSyst}}

\newcommand{\Gammad}{ \ensuremath{ \Gammavalue}}
\newcommand{\GammadStatUncrt}{ \ensuremath{ \GammaStatErr}}
\newcommand{\GammadSystUncrt}{ \ensuremath{ \GammaSystErr}}

\newcommand{\ggratioStatUncrt}{ \ensuremath{ \ratioStatErr}}
\newcommand{\ggratioSystUncrt}{ \ensuremath{ \ratioSystErr}}

\AtlasTitle{Precision measurement of the $B^{0}$ meson lifetime using~$B^{0}\to J/\psi K^{{* 0}}$ decays with the ATLAS detector}

\AtlasAbstract{%
A measurement of the $B^{0}$ meson lifetime %
using \BtoKns\ decays in data from \CoMEnergy\ proton--proton collisions
with an integrated luminosity of \ilumi\ recorded by the ATLAS detector at the LHC is presented.
The measured effective lifetime is
$$
\tau = \BdTauFitVal \pm  \BdTauStatUncrt ~\mathrm{(stat.)} \pm \BdTauSystUncrt ~\mathrm{(syst.)~ps}.
$$
The average decay width, extracted from the effective lifetime and using a parameter from external sources, is
$$
\Gamma_d =  \Gammad \pm  \GammadStatUncrt ~\mathrm{(stat.)} \pm \GammadSystUncrt ~\mathrm{(syst.)} \textrm{~ps}^{-1},
$$
where the uncertainties are statistical and systematic.
The uncertainty due to external sources is negligible at the given precision.
The earlier ATLAS measurement of $\Gamma_s$  in the $\Bso\to J/\psi\phi$ decay was used to derive a value for the ratio of the average decay widths $\Gamma_d$ and $\Gamma_s$ for \Bo\ and \Bso\ mesons respectively,  of
$$
\frac{\Gamma_d }{\Gamma_s } = 0.9910 \pm \ggratioStatUncrt ~\textrm{(stat.)} \pm \ggratioSystUncrt ~\textrm{(syst.)}.
$$
The measured lifetime, average decay width and decay width ratio are in agreement with theoretical predictions and with measurements by other experiments.
This measurement provides the most precise result of the effective lifetime of the $B^{0}$ meson to date.
}

\author{The ATLAS Collaboration}

\AtlasRefCode{ANA-BPHY-2020-09}

\AtlasNote{ANA-BPHY-2020-09-INT1}

\AtlasJournalRef{\EPJC 85 (2025) 736}
\AtlasDOI{10.1140/epjc/s10052-025-14232-8}

\AtlasCoverEgroupAnalysisTeam{atlas-BPHY-2020-09-analysis-team@cern.ch}

\PreprintIdNumber{CERN-EP-2024-279}

\hypersetup{pdftitle={ATLAS document},pdfauthor={The ATLAS Collaboration}}

\begin{document}

\maketitle

\tableofcontents

\newpage

%
%

%
%
%

%

%
\section{Introduction}

The  lifetime, or equivalently the decay width, is one of the fundamental properties of elementary particles, and hence it represents an observable of  primary phenomenological  importance.
In particular, studies of $b$-hadron lifetimes can test our understanding of the weak interaction.
On the theoretical side, lifetimes of $b$-hadrons can be computed systematically in the heavy-quark expansion (HQE)  framework~\cite{Khoze:1983,Shifman:1984wx,Shifman:1986mx,Bigi:1991ir,Bigi:1992su,Bigi:1995jr,Uraltsev:1998bk,Neubert:1997gu}, which permits $b$-hadron observables to be calculated through perturbative expansion in powers of the reciprocal of the $b$-quark mass, $m_b$.
The theoretical models typically provide predictions of the ratios of $b$-hadron lifetimes, since these can be obtained with higher accuracy than is possible for the predictions of absolute lifetimes, due to HQE terms cancelling out in the ratio.
Lifetime differences among the $b$-hadrons are almost entirely attributable to the corrections for Pauli interference and weak annihilation, both of which are enhanced by a relatively large phase space factor where they enter the expansion at the term proportional to $m_{b}^{-3}$ \cite{Lenz:2014jha}. This term also includes the non-perturbative matrix of four-quark operators, which was recently calculated~\cite{Kirk:2017juj} using QCD sum rules~\cite{Shifman:1978bx,Shifman:1978by} formulated in Heavy Quark Effective Theory (HQET)~\cite{Georgi:1990um}. The largest contributions to the total uncertainty in the lifetime ratio, as predicted with the HQE, come from the hadronic matrix elements.

The effective lifetime $\tau_{\Bo}$ is related to the decay widths of the light (L) and heavy (H) mass eigenstates of the
\BoBo\ system via the following equation
taken from Ref.~\cite{Fleischer:2001PhysRevD.63.114015}:
\begin{equation}
\tau_{\Bo} = \frac{1}{\Gamma_d} \frac{1+y^2}{1-y^2},
\label{eq:effective_tau}
\end{equation}
where $\Gamma_d = (\Gamma_\text{L}+\Gamma_\text{H}) / 2$ is the average decay width of the two states,
$y = \Delta \Gamma_d / (2\Gamma_d) = (\Gamma_\text{L}-\Gamma_\text{H}) / (2\Gamma_d)$ is the normalised width difference.

Using the value of $y$ from Ref.~\cite{HFLAV:2022pwe}, inverting \eq(\ref{eq:effective_tau})  %
allows a measured value of $\Gamma_d$ to be extracted.
Its predicted value is
$\Gamma_d = 0.63^{\,+0.11}_{-0.07}~\textrm{ps}^{-1}$~\cite{Lenz:2022rbq}.
Theoretical uncertainties largely cancel out in the decay width ratio $\Gamma_d/\Gamma_s$,
where $\Gamma_s$ is average \Bso\ meson decay width.
\tab \ref{tab:predicted_lifetimes} lists the decay width ratios of \Bo\ and \Bso\ mesons as predicted by the HQE model in Ref.~\cite{Lenz:2022rbq} %
and the lattice QCD model in Ref.~\cite{Becirevic:2001fy}.

\begin{table}[h!]
\begin{center}
\caption{Theoretical predictions for the \Bo~meson to \Bso~meson decay width ratio.\label{tab:predicted_lifetimes} }
\begin{tabular}{l|c}
\hline
Model                           & $\Gamma_d/\Gamma_s$ \\ \hline
HQE \cite{Lenz:2022rbq}         & $1.003 \pm 0.006$   \\
Lattice QCD \cite{Becirevic:2001fy} & $1.00~~ \pm 0.02~~$       \\ \hline
\end{tabular}
\end{center}
\end{table}

Previous measurements of the \Bo\ meson  lifetime have been reported by a variety of collaborations.  %
For the LHC experiments, these are ATLAS~\cite{BPHY-2012-03}, LHCb~\cite{LHCb:2014qsd,LHCb:2014bqh} and CMS~\cite{CMS-BPH-13-008}.
The most recent result comes from Belle~II~\cite{Abudinn2023}.
Other measurements were performed by BaBar~\cite{PhysRevD.73.012004}, CDF~\cite{PhysRevLett.106.121804}, D0~\cite{PhysRevLett.114.062001} and Belle~\cite{Belle:2004hwe}.

This article reports a measurement of the effective lifetime of the \Bo\ meson by the ATLAS experiment.
Data amounting to \ilumi\ of $\sqrt{s}=\CoMEnergy$ proton--proton ($pp$) collisions were collected at the CERN Large Hadron Collider (LHC) during the period 2015--2018.
The \Bo\ is identified using its decay mode \BoToJpsiKnst(892)\ with $J/\psi\to\mu^{+}\mu^{-}$ and $\Kns(892)\to K^{+}\pi^{-}$.
Inclusion of conjugate modes is implied unless noted otherwise.
A combined analysis of the mass and lifetime distributions is performed, where the mass is used to improve discrimination between the signal and background.

The average decay width $\Gamma_d$ and the ratio of widths $\Gamma_d/\Gamma_s$, using the ATLAS-measured
value of $\Gamma_s$ \cite{BPHY-2018-01}, are also determined in this analysis.
The value of $y$ used to extract $\Gamma_d$ is taken from Ref.~\cite{HFLAV:2022pwe}.


%
\section{ATLAS detector, data-taking conditions and simulation}
\label{sec:detector}

The ATLAS detector\footnote{ATLAS uses a right-handed coordinate system
with its origin at the nominal interaction point. The $z$-axis is along
the beam pipe, the $x$-axis points to the centre of the LHC ring, and
the $y$-axis points upward. Cylindrical coordinates $(r,\phi)$ are used
in the transverse plane, $r$ being the distance from the origin and
$\phi$ being the azimuthal angle around the beam pipe. The pseudorapidity
$\eta$ is defined as $\eta = - \ln[\tan (\theta/2)]$ where
$\theta$ is the polar angle.} \cite{PERF-2007-01}
consists of three main components: an inner detector (ID) tracking system
immersed in a $2$~T axial magnetic field, electromagnetic and hadronic calorimeters, and a muon spectrometer (MS).
The ID covers the pseudorapidity range \(|\eta| < 2.5\) and consists of silicon pixel, silicon microstrip, and transition radiation tracking detectors~\cite{ATLAS-TDR-30}. The first point of detection is the insertable B-layer at a radius of 3.3~cm from the beam line; the rest of the pixel detector consists of 92~million pixels capable of measuring the location of a hit to a precision of 10~$\mu$m~\cite{ATLAS-TDR-19,PIX-2018-001}. It is surrounded by the silicon microstrip detector, which consists of 6~million microstrip sensors.
The ID is surrounded by a high-granularity liquid-argon (LAr) sampling electromagnetic calorimeter.
A steel/scintillator tile calorimeter provides hadronic coverage in the central rapidity range. The endcap and forward regions are equipped with LAr calorimeters for electromagnetic and hadronic measurements.
The MS surrounds the calorimeters. It comprises separate trigger and high-precision tracking chambers measuring the deflection of muons in a magnetic field generated by superconducting air-core toroids.
The field integral of the toroids ranges between \num{2.0} and \qty{6.0}{$\textrm{\tesla}\textrm{\metre}$} across most of the detector.
A set of precision chambers covers the region $|\eta| < 2.7$ with three layers of monitored drift tubes, complemented by cathode strip chambers in the forward region, where the background is highest.
The muon trigger system covers the range $|\eta| < 2.4$ with resistive plate chambers in the barrel, and thin gap chambers in the endcap regions.
The luminosity is measured mainly by the LUCID--2~\cite{LUCID2} detector that records Cherenkov light produced in the quartz windows of photomultipliers located close to the beampipe.

Events are selected by the first-level trigger system implemented in custom hardware, followed by selections made by algorithms implemented in software in the high-level trigger~\cite{TRIG-2016-01}. The first-level trigger accepts events from the \qty{40}{\MHz} bunch crossings at a rate below \qty{100}{\kHz}, which the high-level trigger further reduces in order to record complete events to disk at about \qty{1}{\kHz}.

The data were collected during periods with different instantaneous luminosities, so several triggers were used in the analysis~\cite{TRIG-2018-01, TRIG-2019-03}. All triggers were based on the identification of a
\Jpsitomm decay, with various muon transverse momentum (\pT) thresholds (usually $4$~\GeV, $6$~\GeV\ and $11$~\GeV). The composition of the collected data shifts to higher thresholds as the number of $pp$ interactions per bunch crossing (pile-up) increases.
Data quality requirements are imposed on the data, notably on the performance of the MS, ID and calorimeter systems~\cite{DAPR-2018-01}.
The measurement uses $pp$ collision data corresponding to an integrated luminosity of $140.1 \pm 1.2$~fb$^{-1}$~\cite{DAPR-2021-01}, a value
obtained by the LUCID-2 detector~\cite{LUCID2} for the primary luminosity measurements.
An extensive software suite~\cite{SOFT-2022-02} is used in data simulation, in the reconstruction and analysis of real and simulated data, in detector operations, and in the trigger and data acquisition systems of the experiment.



To model the \BoToJpsiKnst\ decay in the ATLAS detector, a Monte Carlo (MC) simulated sample of signal events was generated.
These events were required to have at least one pair of oppositely charged muons with
$\pT > 3.5 ~\GeV$ and $|\eta| < 2.6$  to match the fiducial acceptance of the detector.
Production of $b$-hadrons in $pp$ collisions was simulated using \Pythia[8.244]~\cite{Sjostrand:2014zea}, tuned to ATLAS data
via the A14 set of tuned parameters~\cite{ATL-PHYS-PUB-2014-021}, together with the CTEQ6L1 set of parton distribution functions~\cite{Pumplin:2002vw}.
The MC events were then passed through the ATLAS detector simulation program based on the ATLFAST2 procedure~\cite{SOFT-2010-01} using the \GEANT package~\cite{Geant4}.
The same software as used for processing data from the detector was also used to reconstruct the simulated events.
The MC events were weighted to reproduce the same pile-up,
and the same trigger conditions as in the data,
including the pre-scale factors\footnote{A pre-scale factor defines the fraction of events that are kept from all those that would have passed the trigger requirements.}
applied in data-taking.


%
\section{Event reconstruction and selection}
\label{sec:selection}

Events are first required to pass the trigger selections described in \sect\ref{sec:detector}.
In addition, each event must contain at least one reconstructed primary vertex~\cite{ATL-PHYS-PUB-2015-026},
formed from at least four ID tracks, and at least one pair of oppositely charged muon candidates that are reconstructed using information from the MS and the ID.
The muons used in the analysis are required to meet the \textit{Tight} working point identification criteria~\cite{MUON-2018-03}.
The oppositely charged muon pairs are re-fitted to a common vertex (using ID track parameters only), and the pair is accepted if the quality of the fit meets the requirement \cutChiJpsi \ for its reduced $\chi^2$ value. Here, ndof represents the number of degrees of freedom. Additionally, the $J/\psi$ mass, extracted in the fit, must lie within a range selected to retain 99.7\% of the \Jpsi\ candidates identified in the fits. The size of the chosen range depends on the pseudorapidity of the muons.
The \KstarKPi decay candidates are reconstructed from all tracks that are not identified as muons from the \Jpsi decay.
The oppositely charged tracks are reconstructed under both the $K^{*0}$ and $\Knsbar$ hypotheses,
and the hypothesis giving an invariant mass closer to the world average value for the $K^*$ meson is selected.
Selected tracks satisfying $\pT (K^+)> 1$~\GeV, $\pT (\pi^-)> 0.5$~\GeV\ and $|\eta| < 2.5$ are used,
and the \KstarKPi candidate is required to satisfy $\pT (K^{*0})> 3.5$~\GeV and lie in the range \cutMassKstar.

Each combination of the selected \Jpsitomm\ and \KstarKPi\ candidates is fitted to a common vertex, forming the \BoToJpsiKnst\ candidates.
The vertex fit is constrained by fixing the invariant mass of the two muon tracks to the world average value of the \Jpsipa mass~\cite{PDG}.
The \BoToJpsiKnst\ candidates that fulfil a \cutChiBd\ condition are accepted for further analysis.
The \Bo\ candidate with the lowest $\chi^2 /$ndof is selected in events with more than one candidate meeting the criteria.
In total, \ntotalBdtext\ \Bo\ candidates are selected within the mass range \cutMassBdRange.
This range is chosen so that the sidebands of the mass distribution have enough background events to allow a precise determination of their properties.

The mean number of $pp$ interactions per bunch crossing is 31,
so it is necessary to choose the best candidate for the primary vertex at which the \Bo\ meson is produced. Primary vertex (PV) positions are recalculated after removing any tracks used in the \Bo\ meson reconstruction.
The PV candidate with the smallest three-dimensional impact parameter, $a_0$ (the minimum distance between each PV candidate and the line extrapolated from the reconstructed \Bo\ meson vertex in the direction of the \Bo\ momentum), is used.
For each \Bo\ meson candidate, the proper decay time $t$ is determined using:
\begin{linenomath}
\begin{equation}
t = \frac{ L_{xy}\ m_{{}_{B}} }{ p_{ \mathrm{T}_{B}  }},
\label{TimeFormula}
\end{equation}
\end{linenomath}
where $p_{ \mathrm{T}_{B} }$ is the reconstructed transverse momentum of the \Bo\ meson candidate and $m_{{}_{B}}$ denotes the mass of the \Bo\ meson, taken from Ref.~\cite{PDG}.
The transverse decay length, $L_{xy}$, is the distance in the transverse plane from the primary vertex to the \Bo\ meson decay vertex, projected onto the direction of the \Bo\ transverse momentum.


%
\section{Maximum-likelihood fit}
\label{sec:MLfit}
The lifetime of the \Bo\ meson is extracted from a two-dimensional unbinned maximum-likelihood fit describing both signal and background.
The signal model describes the \BtoKns decay events.
The background model is composed of two contributions.
The first one, referred to as \enquote{prompt} background, corresponds to cases where the \Jpsi meson coming from the \ppJpsiX\ production process is combined with a random \Kns\ candidate in the same event.
The other contribution, referred to as \enquote{combinatorial} background, is formed of events where a \Jpsi meson produced in the decay of any $b$-hadron is combined with a \Kns\ candidate.

The mass and proper decay time are simultaneously fitted using the log-likelihood function:
\begin{alignat}{2} \nonumber %
\ln L = \sum_{i=1}^N w(\tildeti) \ln[ f_{\textrm{sig}} {\cal M}_{\textrm{sig}} (\mathit{m_{i}}) {\cal T}_{\textrm{sig}} ( \mathit{\tildeti}, \sigtildeti, p_{\mathrm{T}_i} )
+ ( 1 - f_{\textrm{sig}} ) {\cal M}_{\textrm{bkg}} (\mathit{m_{i}}) {\cal T}_{\textrm{bkg}} ( \mathit{\tildeti}, \sigtildeti, p_{\mathrm{T}_i} ) ],
\nonumber
\end{alignat}
where $f_{\textrm{sig}}$ is the fraction of signal events in the
total number of events, $N$.
The \BtoKns decay is described by the mass probability density function (PDF) $ {\cal M}_{\textrm{sig}}$ multiplied by the time PDF ${\cal T}_{\textrm{sig}}$.
The prompt and combinatorial backgrounds are modelled by the mass and lifetime PDFs ${\cal M}_{\textrm{bkg}}$ and ${\cal T}_{\textrm{bkg}}$.
The mass $m_{i}$, the proper decay time $\tildeti$, its uncertainty $\sigtildeti$
and the \Bo\ candidate transverse momentum $p_{\mathrm{T_i}}$
are the values measured from the data for each event $i$.
The weight $w(\tildeti)$, as defined by \eq(\ref{eq:LifeTimeWeight}) in \sect\ref{sec:efficiencies},
accounts for the selection efficiency.

\subsection{The invariant mass PDFs}

The ${\cal M}_{\textrm{sig}}$ and ${\cal M}_{\textrm{bkg}}$ PDFs
model the \Bo\ signal and background mass shapes, respectively, in the fitted mass range.
For the signal, the mass is modelled with a Johnson $S_U$-distribution~\cite{Johnson:1949}:
\begin{equation}{\cal M}_{\textrm{sig}} (\mathit{m}_{i}) = \frac{\delta}{ \lambda \sqrt{2\pi} \sqrt{1+\left(\frac{m_i-\mu}{\lambda}\right)^2} }
\exp\left[ - \frac{ 1 }{2 } \left( \gamma + \delta \sinh^{-1} \left(\frac{m_i-\mu}{\lambda}\right) \right)^{2} \right],\nonumber
\label{pdfsig}
\end{equation}
where $\mu$, $\gamma$, $\delta$ and $\lambda$ are free parameters.
For the background, the mass distribution is modelled by the sum of a polynomial and a sigmoid function:
\begin{equation}
\mathcal{M}_{\textrm{bkg}}(m_i) =f_{\textrm{poly}}(1+p_0\cdot m_i) +
(1-f_{\textrm{poly}})\left(1-\frac{s(m_i-m_0)}{\sqrt{1+(s(m_i-m_0))^2}}\right),
\label{pdfbkgs}
\end{equation}
where $f_{\textrm{poly}}$ describes the relative size of the two components and $m_0$, $s$ and $p_0$ are parameters of the PDF.
Here the sigmoid function was added to better describe the background contribution from partially reconstructed $B$ mesons in the lower-mass sideband.

\subsection{The proper decay time PDFs}
The PDF describing the proper decay time for each event $i$ is composed of two terms:
\begin{equation} \label{eq:MLlifetime}
{\cal T}_j ( \tildeti, \sigtildeti, p_{\mathrm{T}_i} )=
P_j ( \tildeti | \sigtildeti, p_{\mathrm{T}_i} ) \cdot C_j (\sigtildeti, p_{\mathrm{T}_i}),
\end{equation}
where $j \in (\textrm{sig, bkg})$ stands for signal or background, respectively.
In each term $P_j$ the function describing the proper decay time behaviour of component $j$ is convolved with the  proper decay time resolution function in order to account for the proper decay time uncertainty. The resolution function is modelled as a sum of three Gaussian distributions:
\begin{equation*}{R} (\tildet'-\tildeti, \sigtildeti) = \sum_{k=1}^{3} f_{\textrm{res}}^{(k)}\frac{1}{ \sqrt{2\pi}~S^{(k)} ~ \sigtildeti }
\exp\left( \frac{ - (\tildet'-\tildeti )^{2} }{2(S^{(k)} \ \sigtildeti )^{2} } \right). \nonumber
\end{equation*}
The scale factors $S^{(k)}$ are free parameters of the fit.  The value of  $\sigtildeti$ is the per-candidate uncertainty in \tildeti, extracted from the vertex fit performed for the  \BoToJpsiKnst\ candidate, described in Section~\ref{sec:selection}.  The parameters $f_{\textrm{res}}^{(k) }$ represent the relative contribution of each Gaussian function and fulfil the normalization condition $\sum f_{\textrm{res}}^{(k)} = 1$.

The signal proper decay time distribution of the \Bo\ signal candidates is modelled as an exponential function
\begin{equation} %
P_{\textrm{sig}} (\tildeti|\sigtildeti, p_{\mathrm{T}_i})= E(\tildet',\tau_{B^0})\otimes R(\tildet'-\tildeti,\sigtildeti),\nonumber
\end{equation}
where ${E(\tildet,\tau_{B^0})= (1/\tau_{B^0})\exp (-\tildet / \tau_{B^0})}$ for $\tildet \ge 0$, with the parameter $\tau_{B^0}$ standing for the \Bo\ lifetime.

The proper decay time PDF for the background candidates, $P_{\textrm{bkg}}$, consists of two parts.
One part accounts for the prompt background and consists of the resolution function $R$ only.
The other part accounts for the combinatorial background and consists of a sum of three exponential functions, each convolved with the resolution function $R$.
In summary, the background proper decay time PDF takes the form:
\begin{alignat}{2}
\label{eq:CombBackgroundLifetimeModel}
P_{\textrm{bkg}} (\tildeti|\sigtildeti, p_{\mathrm{T}_i} ) = \left( f_{\textrm{prompt}}\cdot \delta_{\textrm{Dirac}}(\tildet')
+
(1-f_{\textrm{prompt}}) \sum_{k=1}^{3} b_k \prod_{l=1}^{k-1} (1-b_{l})
E(\tildet',\tau_{\textrm{bkg}_k} )
\right) \otimes R(\tildet'-\tildeti,\sigtildeti).
\end{alignat}
Here the $\tau_{\textrm{bkg}_j}$ are different lifetimes describing three components of the combinatorial background;
the  parameters $b_j$ are the relative fractions of these three background components,
and $f_{\textrm{prompt}}$ is the prompt component's fraction.
Parameters $\tau_{\textrm{bkg}_j}$, $f_{\textrm{prompt}}$ and two of the $b_j$ are free in the fit; $b_3  \equiv 1$ by definition.

The probability terms $C_j (\sigtildeti, p_{\mathrm{T}_i} )$ in \eq(\ref{eq:MLlifetime}) are two-dimensional distributions introduced to describe the difference between signal and background for the per-candidate time uncertainty $\sigtildeti$ and $p_{\mathrm{T}_i}$ values, respectively~\cite{Punzi2004}.
The distributions of  $\sigtildeti$  for signal and background are extracted from the data using the $sPlot$ technique~\cite{Pivk:2004ty} with the \Bo\ candidate mass as the discriminating variable.


%
\section{Efficiencies and corrections}
\label{sec:efficiencies}

The trigger, offline reconstruction, and event selection criteria bias  the reconstructed proper-decay time distribution.
These effects are estimated using signal MC events.
The triggers used in this analysis impose no lower threshold  on either the transverse impact parameter $d_0$ of each muon or on the displacement of the \Jpsi vertex.
The same is true for the offline tracking and event selections.
On the other hand, both the trigger and offline tracking impose an upper limit on $|d_0|$, namely $|d_0| < 10$~mm, for all four final-state tracks from a \BtoKns\ decay.
This results in inefficiency at large values of the proper decay time.

To study inefficiency effects, the signal MC events generated in the fiducial volume of the ATLAS detector,
with no selection applied to the final-state particles from \BtoKns\ decays, are used.
Subsequently, these events are passed through a simulation of the detector response and triggers, followed by offline tracking and vertexing algorithms.
This procedure includes magnetic field simulation, tracking efficiency calibration, and trigger pre-scale emulations to ensure a consistency with  data.
Finally, the event selection criteria listed in \sect\ref{sec:selection} are applied.
The \Bo\ proper decay time distributions obtained before and after the whole chain are used to estimate time inefficiencies for each year of data-taking.
To account for these inefficiencies in the fit, each event $i$ is weighted by a factor $w_i$, which is inversely proportional to the time efficiency function and is defined as:
\begin{equation}
1/w_i(\tildeti) =     p_0 \cdot[ 1 - p_1 \cdot (\textrm{Erf} ((\tildeti-p_3)/p_2) + 1)].
\label{eq:LifeTimeWeight}
\end{equation}
Here \enquote{Erf} denotes the error function, and the values of parameters $p_0$, $p_1$, $p_2$ and $p_3$ are determined in the fit to the MC events.
The fit is applied individually for each year, due to their different trigger and data-taking conditions.
Since the analysis is not sensitive to the absolute value of the efficiency,
only corrections to the shape of the time efficiency functions are applied.
Thus the overall normalisation parameter $p_0$ is chosen to have a weight equal to unity on average for each year of data-taking.
The systematic uncertainties associated with the time efficiency determination and its application in data correction  are explained in the Section \Ref{sec:Syst}  and  included in the systematic uncertainties of the measurement.


%
\section{Systematic uncertainties}
\label{sec:Syst}
The systematic uncertainties in the \Bo\ lifetime arise from several sources.
The uncertainties are estimated from comparisons of nominal and modified fit results and from observed fit biases in modified pseudo-experiments.
The statistical uncertainty of the default fit is $\sigma_\mathrm{stat} = 0.0012~\mathrm{ps}$.
A summary of all sources of uncertainty that contribute to the analysis is presented in \tab \ref{tab:syst_uncertainties},
where the total systematic uncertainty is obtained by adding all of the contributions in quadrature.
Each of the systematic uncertainties is described below.
\begin{itemize}

\item \textbf{Inner detector alignment:}
Effects due to the ATLAS ID misalignment are dominated by the global length scale biases. These are characterised by detector geometry distortions along the track trajectory and affect  both the radial and longitudinal directions, causing  biases to the measured transverse and longitudinal momenta ~\cite{IDTR-2019-05}.
This effect manifests itself as a shift in the reconstructed invariant mass of known resonances.
ATLAS determined that the  global length  bias  equally affects both radius and length of detector leading to  momenta biases: $p_\mathrm{T}^{\prime} =p_\mathrm{T}(1+\epsilon_\mathrm{s})$  and $p_\mathrm{z}^{\prime} =p_\mathrm{z}(1+\epsilon_\mathrm{s})$, where  $\epsilon_\mathrm{s}$  is a scale factor.
On the basis of  \Jpsitomm events the scale $\epsilon_s = -0.085\%$ has been determined. The value is  consistent  with the one  extracted for  Z~\ra \mumu in the same paper ~\cite{IDTR-2019-05}.

In this analysis the tracks forming the \Bo\ candidate vertex are re-fitted with the \Jpsi\ candidate mass constrained to the world average value, effectively removing the misalignment effect from the data.
The impact of the ID misalignment in the lifetime determination is estimated by performing an alternative fit, where the tracks forming the $B^0$ meson vertex are re-fitted without constraining the \Jpsi\ mass to its world average value.
It is worth noting that the measurement of the momentum bias, using invariant mass of resonances ~\cite{IDTR-2019-05}, involving radial and longitudinal scale length biases,  also includes momentum biases  due to magnetic field distortions.
The proper decay time in this analysis is calculated using the ratio of $ L_{xy}$ and $p_{ \mathrm{T}_{B} }$ with \eq(\Ref{TimeFormula}). The value of $ L_{xy}$ is sensitive to the radial length scale bias in the same way as $p_{ \mathrm{T}_{B} }$, while $ L_{xy}$ is not sensitive to the magnetic field distortions.
So in the proper decay time, the misalignment effects only partially cancel.
To follow a conservative approach, the alternative fit without a \Jpsi\ mass constraint was performed and compared to the default fit that includes \Jpsi\ constraint, resulting in  a difference   $0.9\,\sigma_\mathrm{stat}$ of the \Bo\ lifetimes,  which is included into systematic uncertainty due to misalignment. Additionally, to account for the momentum scale bias affecting low-$p_\mathrm{T}$ hadrons, the $p_\mathrm{T}$ values of hadrons from the $K^{*0}$ decay are altered by $-0.085\%$~\cite{IDTR-2019-05}. The difference of lifetimes between this alternative fit and the default one is conservatively taken as a systematic uncertainty associated with the momentum bias of the low-$p_\mathrm{T}$ hadrons due to ID misalignment.
The two effects have been symmetrised and summed in quadrature, giving the value of $1.0\,\sigma_\mathrm{stat}$, that serves as an estimate for the total systematic uncertainty due to ID misalignment.

\item \textbf{Choice of mass window:}
The sensitivity of the lifetime fit to the chosen \Bo\ mass window is estimated by choosing other intervals
as alternatives to the default mass range of 5000--5650~\MeV.
Several other intervals are studied, varying the mass range limits  by  $\pm$20~MeV or   $\pm$40~MeV.
The largest deviation of the alternative fit lifetimes from the default value, $0.9\,\sigma_\mathrm{stat}$, is taken as an estimate of this systematic uncertainty.

\item \textbf{Choice of primary vertex:}
As mentioned in Section \Ref{sec:selection}, among the multiple primary vertices typically found in an event, the PV with the smallest $a_0$ is chosen by default.
An alternative approach is to use the PV with the highest sum of the squares of the constituent tracks' transverse momenta, $\sum p_\textrm{T}^2$.
The difference between the \Bo\ fitted lifetimes obtained using the default and alternative approaches is
negligibly small compared to the statistical uncertainty of the fit. %
The small difference is due to the fact that
the PV position resolution in the transverse plane is same as the beam-spot size.
A PV position difference in the $z$-direction is not relevant,
since the lifetime is extracted in the transverse plane, as defined in \eq(\ref{TimeFormula}).

\item \textbf{Time efficiency:} Time efficiency curves are used to correct for the proper decay time dependence of reconstruction, trigger and candidate selection  inefficiencies.
They are built with information from MC events, as is described in \sect\ref{sec:efficiencies}.
Two alternative time efficiency functions are obtained by replacing the error function in \eq(\ref{eq:LifeTimeWeight}) with either a hyperbolic tangent function or $(x^2+1)^{-1/2}$.
The larger of the changes in the fitted lifetime, of size $0.6\,\sigma_\mathrm{stat}$,  is taken as a systematic uncertainty.
To account for potential systematic effects due to the limited MC sample size,
a large number of alternative time efficiency functions are obtained by smearing the number of MC events in the time bins used to determine the time efficiency function.
These functions are then used to rerun the unbinned maximum-likelihood fit on the data.
The set of these fit results is characterized by its mean value and standard deviation, and these contribute a systematic uncertainty of $0.8\,\sigma_\mathrm{stat}$.
An alternative fit using data events satisfying the requirement $t<8~\mathrm{ps}$, where the efficiency is almost constant, is performed to validate
the modelling of the efficiency dependence on the lifetime for high lifetimes, where the efficiency decrease is large.
This systematic test examines associated effects, such as modelling of tracking efficiencies, triggers, variations in the magnetic field, and provides a conservative estimate of their uncertainties.
The systematic uncertainty derived from this test is $0.5\,\sigma_\mathrm{stat}$. Choosing a somewhat different upper requirement on $t$ has minimal impact on the test results.
Another test is performed by building alternative time efficiency functions based on a  signal MC with  helicity angles shaped according to full PDF of   \BoToJpsiKnst \ measured by LHCb \cite{PhysRevD.88.052002}. The resulting lifetime is identical with the default one within $0.3\,\sigma_\mathrm{stat}$.
All  contributions to the systematic uncertainty associated with the time efficiency are added in quadrature.

\item \textbf{Best-candidate selection:}
90\% of events that meet all the selection criteria (listed in \sect\ref{sec:selection}) contain just one $B^0$ meson candidate.
The others have 2.1 candidates on average.
These were all found to share the same \Jpsi\ in their event. Only the candidate with the lowest $\chi^2$/ndof in the event is retained for the analysis.
The systematic uncertainty due to candidate selection is estimated by applying an alternative time efficiency curve calculated from MC events by using the true generated candidates instead of the lowest $\chi^2$/ndof candidate. The systematic uncertainty estimated from this test is $0.3\,\sigma_\mathrm{stat}$.
Both efficiency functions, based on different candidate selection criteria, were validated on signal MC events, reproducing the lifetime value with a precision well below the statistical uncertainty.

\item \textbf{Mass fit model:} To estimate how mismodelling of the mass distribution affects the fitted \Bo\ lifetime, three contributions are considered.
The first contribution comes from kinematic reflections, where one of the tracks is either misidentified as another particle or completely missing.
The MC simulation is used to model the shapes of the kinematic reflections, and
four channels that are non negligible,  \BpToJpsiK,    $\Bs \to J/\psi \bar{K}^{*0}(K^-\pi^+)$,  $\Bs \to J/\psi K^+K^-$ and $\Bo \to J/\psi \rho (\pi^+ \pi^-)$ (with intermediate $\phi$ and $\rho^0$ resonances accounted for in the last two channels), are included in the fit to data with freely floating relative fractions, while the lifetimes are shaped same as the combinatorial background. Contributions of the various channels are included either individually or together. These alternative fits result in the $\Bo$ lifetime values that differ from the nominal one by up to $1.0\,\sigma_\mathrm{stat}$.
For the $\Bs \to J/\psi \bar{K}^{*0}(K^-\pi^+)$ decay contribution, the signal-like lifetime model was also tested by fixing the lifetime to the $\Bs$ flavor-specific world average value~\cite{PDG}, producing results consistent with the fit that modeled the lifetime distribution as the combinatorial background across all four channels.
Other channels like $\Lambda_b \rightarrow J/\psi K p$,  $\Lambda_b \rightarrow J/\psi \Lambda^0$ and $\Lambda_b \rightarrow J/\psi p \pi^-$ are found to be negligible.
The resulting deviation from the default fit's \Bo\ lifetime value is used as a systematic uncertainty.
Secondly, the signal mass is tested by running alternative fits with different signal mass models, where the Johnson $S_U$-distribution in \eq(\ref{pdfsig}) is replaced by a double-sided Crystal Ball distribution~\cite{DSCB} or Student's $t$-distribution~\cite{studentT}.
The maximum deviation from the default fit is included in the systematic uncertainty.
Lastly,
the systematic uncertainty from the choice of background mass PDF
is estimated by using alternative functions
where the sigmoid component
of \eq(\ref{pdfbkgs})
is replaced by either an $\arctan$ function
or $(1+|s(m_i - m_0)|)^{-1}$.
Systematic deviations from all the variations described above, added in quadrature give a systematic error of $1.3\,\sigma_\mathrm{stat}$ and they are included in the total systematic uncertainty.

\item \textbf{Mass-time correlation:}
The correlations between the invariant mass and the pseudo-proper decay time, and their potential impact on the fit results, is studied.
For the signal events the MC simulation shows a negligible correlation between these two variables.
A data driven method is used for background events, the mass-sideband regions $5.000~\GeV< m_B <5.100~\GeV$ (left) and $5.450~\GeV<m_B <5.650~\GeV$ (right) are divided into six bins of equal size of 50 MeV.
Then using the lifetime background description of the default PDF function, \eq(\Ref{eq:CombBackgroundLifetimeModel}), the component fractions  $ f_{\textrm{prompt}}$,  $b_{1}$, and $b_{2}$
are determined through likelihood fits for each of the six mass bins. These values are found to be linearly dependent on the mass.
An alternative PDF is then  constructed  in the mass-lifetime fit with $f_{\textrm{prompt}}(m_i) = a + b(m_i-5.279~\GeV)$, $b_{1}(m_i) = c_{1} + d_{1}(m_i-5.279~\GeV)$ and $b_{2}(m_i) = c_{2} + d_{2}(m_i-5.279~\GeV)$, where $a$, $b$, $c_{1}$,  $d_{1}$, $c_{2}$ and  $d_{2}$ are free parameters of the fit.
These are found to be consistent with the linear dependency observed in the mass-sideband bins.
This  alternative fit results in the $B^0$ lifetime value, that is  by $1.9\,\sigma_\mathrm{stat}$ different from the default one. This difference is taken as a systematic uncertainty of the fit due to mass-time correlation.

\item \textbf{Proper decay time fit model:}
The proper decay time resolution model is a sum of three Gaussian distributions with a common mean at zero and freely floating relative fractions. Alternative resolution models employing two or four Gaussian functions were evaluated.
The impact of choosing the three-Gaussian resolution model instead of an alternative one is treated as a systematic uncertainty.

The decay time probability density function for background events is constructed by combining three exponential functions. The use of this model is validated by examining events gathered from the invariant mass sidebands. An alternative model with four exponential functions was tested, resulting in negligible systematic uncertainty.

\item \textbf{Conditional probability model:}
The probability terms $C_j (\sigma_{\tildeti}, p_{\mathrm{T}_i} )$ in \eq(\ref{eq:MLlifetime}) are two-dimensional distributions
introduced to describe differences between signal and background events with regard to their per-candidate time uncertainties $\sigma_{\tildeti}$ and $p_{\mathrm{T}_{i}}$ values. One source of systematic uncertainty is the binning choice. Another source is the choice of method used to smooth the binned distribution. The default model uses a kernel smoothing method based on Gaussian smearing with variable width to cope with zero-content bins, and additional interpolation between bins is performed as a systematic test. The systematic uncertainty in the \Bo\ lifetime from these sources is evaluated as the maximum deviation from the nominal fit result when combining the variations described above, and is found to be $0.5\,\sigma_\mathrm{stat}$. An additional test was conducted to assess the impact of using Punzi distributions derived from the same dataset, where the dataset was split into halves and a Punzi distribution was extracted from the second half. The observed difference of $0.3\,\sigma_\mathrm{stat}$ between the fit results using the Punzi distribution from the original dataset and the one from the second half is incorporated into the total systematic uncertainty in quadrature.

\item \textbf{Fit model test with pseudo-experiments:}
The fit model can be sensitive to some of the nuisance parameters (e.g.\ parameters modelling the mass and background lifetime distribution shapes), which could potentially lead to a bias in the measured physics parameters, even if the model describes the fitted data well.
The nominal fit model, with its parameter values extracted from the fit to data, was used to generate an ensemble of pseudo-experiments, which were subsequently fitted with the same model. The mean deviation of the \Bo\ lifetime values extracted from the pseudo-experiment fits is assigned as a systematic uncertainty.

\end{itemize}

Various tests are done to confirm the stability of the results across trigger selections and $\Bo$ candidate $p_\textrm{T}$ range.
The data sample is divided according to three categories of triggers, based on the muon $p_\textrm{T}$ thresholds used in the trigger algorithms. Additionally, three subsamples are created by splitting into three $p_\textrm{T}$ ranges with similar data statistics. In both cases, the resulting $\Bo$ lifetime value remains consistent with the full dataset fit.

The systematic uncertainties listed in \tab\ref{tab:syst_uncertainties} are treated as uncorrelated and thus added in quadrature.
\begin{table}[h]
\caption{Summary of systematic uncertainties assigned to the value of the \Bo\ lifetime.}
\label{tab:syst_uncertainties}

\centering \renewcommand{\arraystretch}{1.2}
\sisetup{round-mode=places,
round-precision=5,
round-minimum=0.00001,
table-format=<1.5}
\begin{tabular}{lc}
\hline\hline
Source of uncertainty     & {Systematic uncertainty $[\textrm{ps}]$}    \\
\hline
ID alignment           &     \phantom{<} \num{\tauBdSystToTableA}    \\
Choice of mass window     &     \phantom{<} \num{\tauBdSystToTableB}    \\
Time efficiency           &     \phantom{<} \num{\tauBdSystToTableD}    \\
Best-candidate selection    &                  \phantom{<} \num{\tauBdSystToTableK}\\
Mass fit model            &     \phantom{<} \num{\tauBdSystToTableE}    \\
Mass-time correlation     &     \phantom{<} \num{\tauBdSystToTableL}    \\
Proper decay time fit model            &     \phantom{<} \num{\tauBdSystToTableH}    \\
Conditional probability model            &     \phantom{<} \num{\tauBdSystToTableG}    \\
Fit model test with pseudo-experiments &     \phantom{<} \num{\tauBdSystToTableF}    \\
\hline
Total            &     \phantom{<} \BdTauSystUncrt~~              \\
\hline\hline
\end{tabular}
\end{table}

\newpage


%
\section{Results}
\subsection{\Bo\ lifetime result}
\begin{figure}[htbp]
\begin{center}
\subfloat[]{
\includegraphics[width=0.48\textwidth]{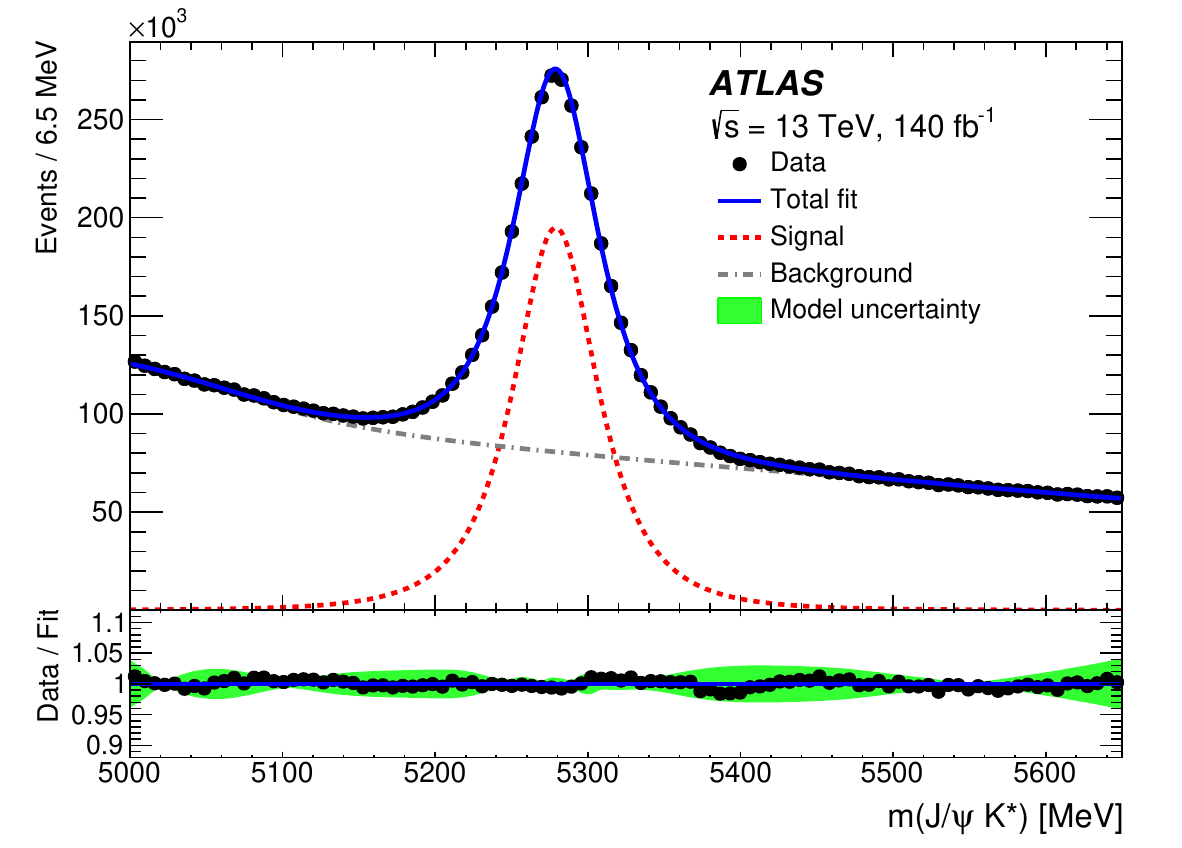}
}\\
\subfloat[]{
\includegraphics[width=0.48\textwidth]{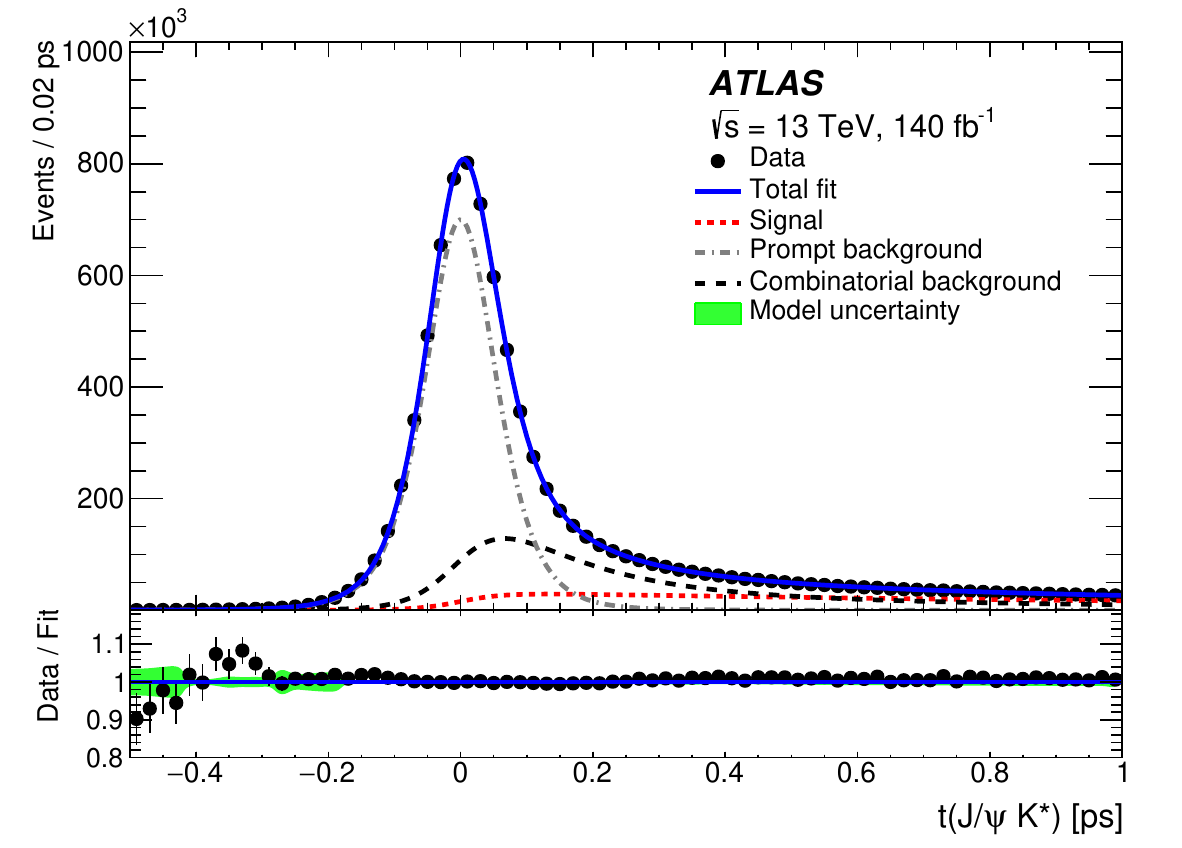}
}
\subfloat[]{
\includegraphics[width=0.48\textwidth]{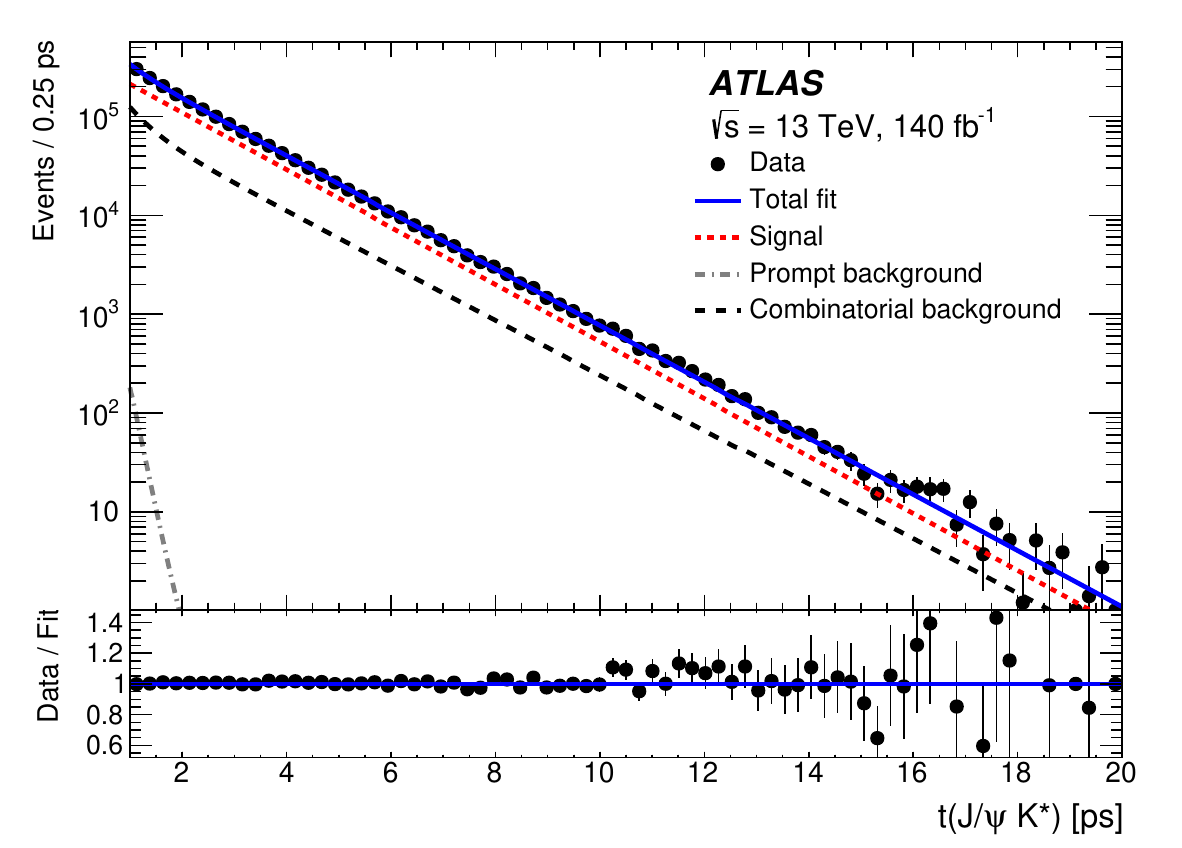}
}
\caption{(a) Invariant mass fit projection and proper decay time fit projection for the \BoToJpsiKnst\ sample shown in two different proper decay time ranges: (b) $\tildet \in (-0.5;1.0)$~ps and (c) $\tildet \in (1;20)$~ps. The solid blue line shows the total fit while the short-dashed red line shows the signal. The sum of prompt and combinatorial backgrounds in the mass fit projection is represented by the dash-dotted grey line. The lifetime projections show the prompt background as a dash-dotted grey line and the combinatorial background as a long-dashed black line. The lower panel of each figure shows the ratio of each data point to the total fitted value. The green band in the ratio plot represents the envelope of model variations included in the systematic uncertainty, while the bars on the data points indicate statistical uncertainties. In subfigure (c), the model variation band is too small to be visible. }
\label{fig:tau_projections}
\end{center}
\end{figure}

The \Bo\ effective lifetime value measured with a total of $2\,450\,500\pm 2400$ \Bo\ signal events\footnote{
The number of signal events is
calculated from the total number of event candidates and the signal fraction value of $0.23175 \pm 0.00022$ obtained by the fit.}
is found to be
\begin{equation}
\tau_{\Bo} = \BdTauFitVal \pm \BdTauStatUncrt ~\textrm{(stat.)} \pm \BdTauSystUncrt ~\textrm{(syst.)}~\textrm{ps}.
\label{eq:tau_result}
\end{equation}
The invariant mass and lifetime projections of the two-dimensional maximum-likelihood fit are shown in \fig\ref{fig:tau_projections}.
The lower panels show the ratio of each data point to the fitted value and, for the top figure only, a band that represents the envelope of mass fit model variations included in the systematic uncertainty.

\begin{figure}[htbp]
\begin{center}
\includegraphics[width=0.55\textwidth]{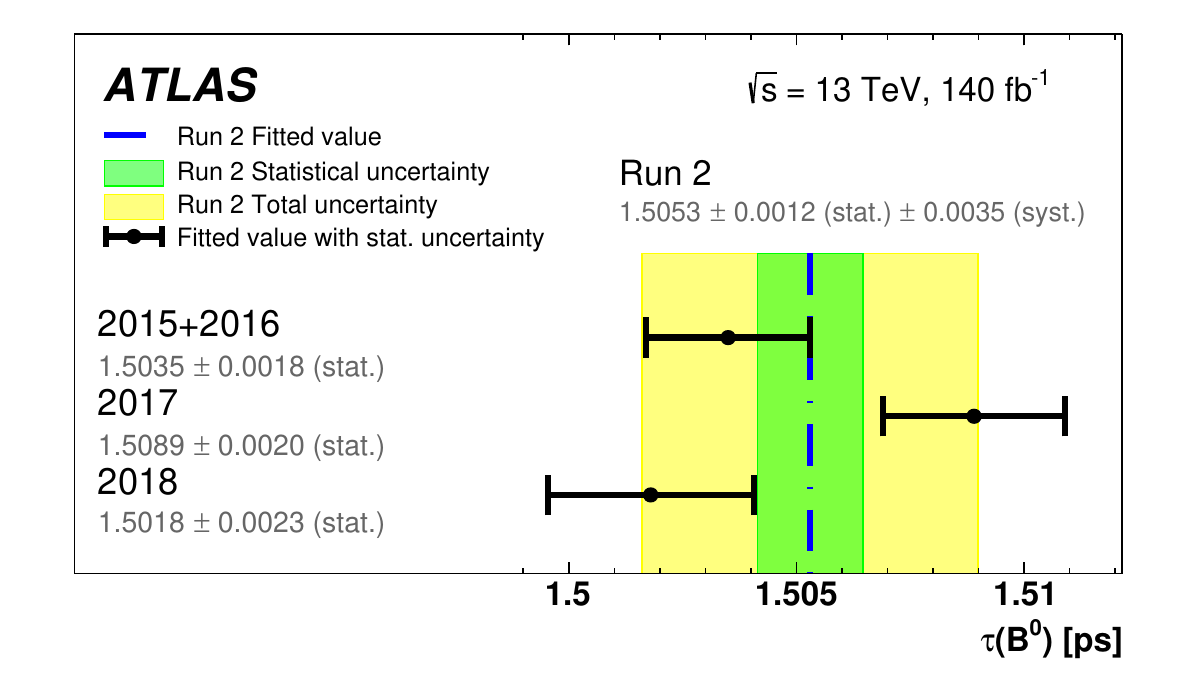}
\caption{The fitted values of the \Bo\ lifetime, measured with \BoToJpsiKnst\ decays, for the 2015+2016, 2017 and 2018 subsamples compared to the value for the whole sample. The \Bo\ lifetime value for each subsample is shown by a black point, with the error bar indicating the statistical uncertainty.}
\label{fig:tau_year_stability}
\end{center}
\end{figure}

As a consistency and stability test, the \Bo\ lifetime value was fitted separately for each data-taking period (2015+2016, 2017 and 2018). Figure~\ref{fig:tau_year_stability} shows the degree of stability over time. The $p$-value for consistency of the three individual results, accounting for statistical uncertainties only, is 0.038.

\subsection{Determination of the \Bo\ average decay width $\Gamma_d$ and the ratio $\Gamma_d/\Gamma_s$}

Equation~({\ref{eq:effective_tau}}) is used to extract the average decay width $\Gamma_d$ from the measured effective lifetime $\tau_{\Bo}$.
The value of $2y = \Delta\Gamma_d /\Gamma_d = 0.001\pm0.010$ is taken from the combination of measurements by HFLAV~\cite{HFLAV:2022pwe}.
Using the value of $y$ and the fitted \Bo\ lifetime, the average decay width is
\begin{equation*}
\Gamma_d = \Gammad \pm \GammadStatUncrt ~\textrm{(stat.)} \pm \GammadSystUncrt~\textrm{(syst.)} ~ \textrm{ps}^{-1}.
\end{equation*}
At the given precision the uncertainty due to the external parameter $y$ is negligible.

The ratio of the average decay widths of the \Bo\ and \Bs\ mesons,
$\Gamma_d/\Gamma_s$, is another quantity of interest in this analysis.
The average decay width $\Gamma_s$ was measured previously by the ATLAS Collaboration~\cite{BPHY-2018-01} as
$\Gamma_s = 0.6703 \pm 0.0014~\textrm{(stat.)} \pm 0.0018~\textrm{(syst.)}~\textrm{ps}^{-1}$.
The resulting ratio is
\begin{equation*}
\frac{\Gamma_d }{\Gamma_s } = 0.9910 \pm \ggratioStatUncrt ~\textrm{(stat.)} \pm \ggratioSystUncrt ~\textrm{(syst.)},
\end{equation*}
where the statistical and systematic uncertainties are propagated from the quantities above.
In the ratio $\Gamma_d/\Gamma_s$ the systematic uncertainties of the ATLAS measurements of $\tau_{\Bo}$ and $\Gamma_s$ primarily come from different sources. They are therefore treated as uncorrelated and can simply be added in quadrature.
\subsection{Comparison with other measurements and theory predictions}
\begin{figure}[htbp]
\begin{center}
\includegraphics[width=0.80\textwidth]{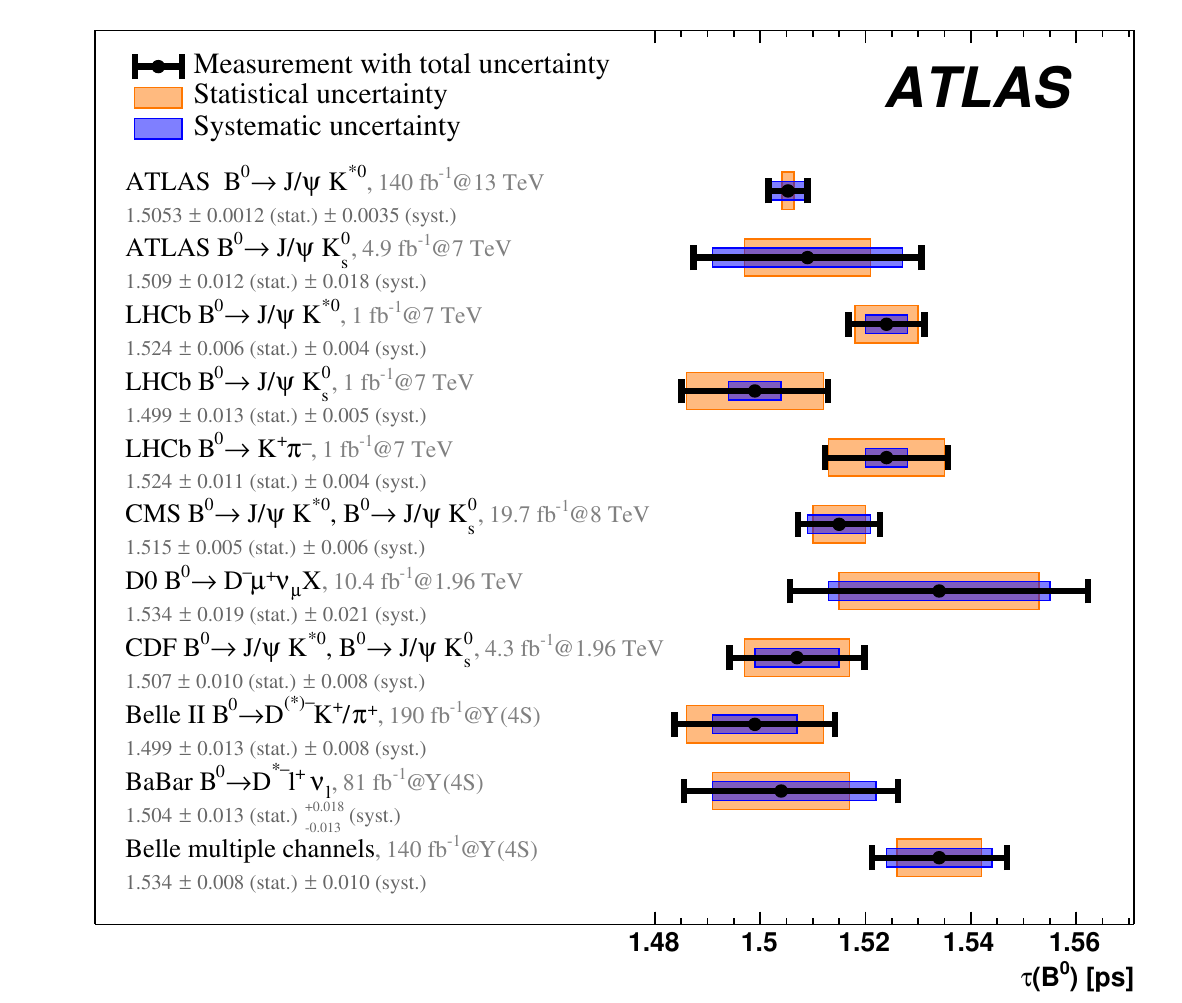}
\caption{
A comparison of the current ATLAS result for the \Bo\ lifetime with the previous ATLAS result~\cite{BPHY-2012-03} in the \BtoKs\ channel, and with those from other experiments.
For the LHC, there are two measurements by LHCb~\cite{LHCb:2014qsd} in \BoToJpsiKnst\ and \BtoKs\ decays,
another LHCb result~\cite{LHCb:2014bqh} in the $\Bo \ra K^+ \pi^-$ channel,
and the CMS~\cite{CMS-BPH-13-008} combined result for \BoToJpsiKnst\ and \BtoKs\ decays.
Other contributions come from the Tevatron experiments: from D0~\cite{PhysRevLett.114.062001} in the $\Bo \ra D^{-}\mu^{+} \nu_{\mu}$ channel, and from CDF~\cite{PhysRevLett.106.121804} with a combined result for \BoToJpsiKnst\ and \BtoKs.
From $e^+e^-$ colliders, the leading result from Belle II~\cite{Abudinn2023} in the $\Bo \ra D^{(*)-} K^+ / \pi^+$ channel and the result from BaBar~\cite{PhysRevD.73.012004} in the $B^{0} \ra D^{*-}\ell^{+}\nu_{\ell}$ channel are shown. The last measurement presented is a combination of multiple channels, performed by the Belle experiment~\cite{Belle:2004hwe}.
This combination includes measurements from the following decays: $\Bo \ra D^{*-} \ell^+ \nu$, $\Bo \ra D^{*-} \pi^+$, $\Bo \ra D^- \pi^+$, $\Bo \ra D^{*-} \rho^+$, $\Bo \ra J/\psi K^{*0}$, $\BtoKs$.
}
\label{fig:tau_Comparisons}
\end{center}
\end{figure}

A comparison of the current \Bo\ lifetime result with the latest and most precise results from other experiments is shown in \fig\ref{fig:tau_Comparisons}.
This result is compatible with most of the other measurements.
In the case of LHCb, three precise measurements are quoted,  each of them in a different channel.
The one in the \BtoKs\ channel agrees with the ATLAS result.
In each of the following comparisons, the quantity $\sigma$ is estimated by combining the uncertainties from ATLAS and the other source.
Thus, the other two LHCb results have lifetimes larger than the current ATLAS value by  $\tauDifftoLHCbKsta\sigma$  in \BtoKns\ and $\tauDifftoLHCbKpi\sigma$  in $\Bo\to K^{+}\pi^{-}$.
The latest world average measured value, published in 2024, is $1.517 \pm 0.004~\textrm{ps}$~\cite{PDG},
which differs from this measurement by $ \tauDifftoHflav\sigma$.
This value includes all results obtained in the history of \Bo\ lifetime measurements.

Compared to the previous ATLAS measurement~\cite{BPHY-2012-03} at $\sqrt{s} = \CoMEnergySeven$, the new ATLAS measurement reduces the systematic uncertainty by a factor of 4.7. Although the previous measurement was based on \BtoKs, which typically has larger systematic uncertainties because of the longer-lived \KzeroS,  much of the improvement comes from the insertable B-layer installed in ATLAS after \RunOne, which provides more precise vertexing and hence better time resolution. Furthermore, better ID alignment in \RunTwo than in \RunOne allowed the related systematic uncertainties to be reduced. In addition, the larger dataset recorded in \RunTwo allowed more precise modelling of the decays and a decrease in related uncertainties.

The  ATLAS $\Gamma_d$ value extracted from this \Bo\ lifetime measurement is compatible with the HQE theory prediction of $0.63^{\,+0.11}_{-0.07}~\textrm{ps}^{-1}$~\cite{Lenz:2022rbq} within $\GammaDifftoHQE\sigma$, where the  uncertainty $\sigma$ is estimated by combining the uncertainty of this result and the uncertainty given in Ref.~\cite{Lenz:2022rbq}.
The ATLAS result for $\Gamma_d/\Gamma_s$  is compatible with the theory predictions of HQE and lattice QCD models~\cite{Lenz:2022rbq,Becirevic:2001fy}
(see also \tab\ref{tab:predicted_lifetimes})
within $\RatioDiffHQE\sigma$  and  $\RatioDiffLatice\sigma$, respectively, and with the experimental average ($1.001 \pm 0.004$)~\cite{HFLAV:2022pwe} within $1.7\sigma$, where the respective $\sigma$ values are estimated by combining the uncertainties from the ATLAS measurements and the other sources.

\FloatBarrier
%


%
\section{Conclusions}
This paper presents a measurement of the \Bo\ effective lifetime and the average decay width $\Gamma_d$ using \BoToJpsiKnst\ events reconstructed from a \ilumi\ data sample of $pp$ collisions collected with the ATLAS detector during the $\sqrt{s}=\CoMEnergy $ LHC run.
The \Bo\ effective lifetime is measured to be
$\tau_{\Bo} = \BdTauFitVal\ \pm\ \BdTauStatUncrt ~\textrm{(stat.)}\ \pm\ \BdTauSystUncrt ~\textrm{(syst.)} ~ \textrm{ps}$.
This ATLAS result is compatible with other experimental measurements and is the most precise measurement to date.
The measured average decay width of the heavy and light \Bo\ mass eigenstates is
$\Gamma_d = \Gammad \pm \GammadStatUncrt ~\textrm{(stat.)} \pm \GammadSystUncrt ~\textrm{(syst.)} ~ \textrm{ps}^{-1}$.
This value is in good agreement with the theory prediction. %
The measured average decay width $\Gamma_d$ is combined with the average decay width $\Gamma_s$ measured previously by ATLAS %
to obtain the ratio $\Gamma_d/\Gamma_s = 0.9910   \pm \ggratioStatUncrt ~\textrm{(stat.)} \pm \ggratioSystUncrt ~\textrm{(syst.)}$.
This result is compatible with the theory predictions from HQE and lattice QCD calculations,
as well as with the experimental average. %

\newpage


\section*{Acknowledgements}
%

%

%
%

%
%

We thank CERN for the very successful operation of the LHC and its injectors, as well as the support staff at
CERN and at our institutions worldwide without whom ATLAS could not be operated efficiently.

The crucial computing support from all WLCG partners is acknowledged gratefully, in particular from CERN, the ATLAS Tier-1 facilities at TRIUMF/SFU (Canada), NDGF (Denmark, Norway, Sweden), CC-IN2P3 (France), KIT/GridKA (Germany), INFN-CNAF (Italy), NL-T1 (Netherlands), PIC (Spain), RAL (UK) and BNL (USA), the Tier-2 facilities worldwide and large non-WLCG resource providers. Major contributors of computing resources are listed in Ref.~\cite{ATL-SOFT-PUB-2025-001}.

We gratefully acknowledge the support of ANPCyT, Argentina; YerPhI, Armenia; ARC, Australia; BMWFW and FWF, Austria; ANAS, Azerbaijan; CNPq and FAPESP, Brazil; NSERC, NRC and CFI, Canada; CERN; ANID, Chile; CAS, MOST and NSFC, China; Minciencias, Colombia; MEYS CR, Czech Republic; DNRF and DNSRC, Denmark; IN2P3-CNRS and CEA-DRF/IRFU, France; SRNSFG, Georgia; BMBF, HGF and MPG, Germany; GSRI, Greece; RGC and Hong Kong SAR, China; ISF and Benoziyo Center, Israel; INFN, Italy; MEXT and JSPS, Japan; CNRST, Morocco; NWO, Netherlands; RCN, Norway; MNiSW, Poland; FCT, Portugal; MNE/IFA, Romania; MESTD, Serbia; MSSR, Slovakia; ARIS and MVZI, Slovenia; DSI/NRF, South Africa; MICIU/AEI, Spain; SRC and Wallenberg Foundation, Sweden; SERI, SNSF and Cantons of Bern and Geneva, Switzerland; NSTC, Taipei; TENMAK, T\"urkiye; STFC/UKRI, United Kingdom; DOE and NSF, United States of America.

Individual groups and members have received support from BCKDF, CANARIE, CRC and DRAC, Canada; CERN-CZ, FORTE and PRIMUS, Czech Republic; COST, ERC, ERDF, Horizon 2020, ICSC-NextGenerationEU and Marie Sk{\l}odowska-Curie Actions, European Union; Investissements d'Avenir Labex, Investissements d'Avenir Idex and ANR, France; DFG and AvH Foundation, Germany; Herakleitos, Thales and Aristeia programmes co-financed by EU-ESF and the Greek NSRF, Greece; BSF-NSF and MINERVA, Israel; NCN and NAWA, Poland; La Caixa Banking Foundation, CERCA Programme Generalitat de Catalunya and PROMETEO and GenT Programmes Generalitat Valenciana, Spain; G\"{o}ran Gustafssons Stiftelse, Sweden; The Royal Society and Leverhulme Trust, United Kingdom.

In addition, individual members wish to acknowledge support from Armenia: Yerevan Physics Institute (FAPERJ); CERN: European Organization for Nuclear Research (CERN PJAS); Chile: Agencia Nacional de Investigaci\'on y Desarrollo (FONDECYT 1230812, FONDECYT 1230987, FONDECYT 1240864); China: Chinese Ministry of Science and Technology (MOST-2023YFA1605700, MOST-2023YFA1609300), National Natural Science Foundation of China (NSFC - 12175119, NSFC 12275265, NSFC-12075060); Czech Republic: Czech Science Foundation (GACR - 24-11373S), Ministry of Education Youth and Sports (ERC-CZ-LL2327, FORTE CZ.02.01.01/00/22\_008/0004632), PRIMUS Research Programme (PRIMUS/21/SCI/017); EU: H2020 European Research Council (ERC - 101002463); European Union: European Research Council (ERC - 948254, ERC 101089007), Horizon 2020 Framework Programme (MUCCA - CHIST-ERA-19-XAI-00), European Union, Future Artificial Intelligence Research (FAIR-NextGenerationEU PE00000013), Italian Center for High Performance Computing, Big Data and Quantum Computing (ICSC, NextGenerationEU); France: Agence Nationale de la Recherche (ANR-20-CE31-0013, ANR-21-CE31-0013, ANR-21-CE31-0022, ANR-22-EDIR-0002), Investissements d'Avenir Labex (ANR-11-LABX-0012); Germany: Baden-Württemberg Stiftung (BW Stiftung-Postdoc Eliteprogramme), Deutsche Forschungsgemeinschaft (DFG - 469666862, DFG - CR 312/5-2); Italy: Istituto Nazionale di Fisica Nucleare (ICSC, NextGenerationEU), Ministero dell'Università e della Ricerca (NextGenEU PRIN20223N7F8K M4C2.1.1); Japan: Japan Society for the Promotion of Science (JSPS KAKENHI JP22H01227, JSPS KAKENHI JP22H04944, JSPS KAKENHI JP22KK0227, JSPS KAKENHI JP23KK0245); Netherlands: Netherlands Organisation for Scientific Research (NWO Veni 2020 - VI.Veni.202.179); Norway: Research Council of Norway (RCN-314472); Poland: Ministry of Science and Higher Education (IDUB AGH, POB8, D4 no 9722), Polish National Agency for Academic Exchange (PPN/PPO/2020/1/00002/U/00001), Polish National Science Centre (NCN 2021/42/E/ST2/00350, NCN OPUS nr 2022/47/B/ST2/03059, NCN UMO-2019/34/E/ST2/00393, NCN \& H2020 MSCA 945339, UMO-2020/37/B/ST2/01043, UMO-2021/40/C/ST2/00187, UMO-2022/47/O/ST2/00148, UMO-2023/49/B/ST2/04085); Slovenia: Slovenian Research Agency (ARIS grant J1-3010); Spain: Generalitat Valenciana (Artemisa, FEDER, IDIFEDER/2018/048), Ministry of Science and Innovation (MCIN \& NextGenEU PCI2022-135018-2, MICIN \& FEDER PID2021-125273NB, RYC2019-028510-I, RYC2020-030254-I, RYC2021-031273-I, RYC2022-038164-I), PROMETEO and GenT Programmes Generalitat Valenciana (CIDEGENT/2019/027); Sweden: Carl Trygger Foundation (Carl Trygger Foundation CTS 22:2312), Swedish Research Council (Swedish Research Council 2023-04654, VR 2018-00482, VR 2021-03651, VR 2022-03845, VR 2022-04683, VR 2023-03403), Knut and Alice Wallenberg Foundation (KAW 2018.0157, KAW 2018.0458, KAW 2019.0447, KAW 2022.0358); Switzerland: Swiss National Science Foundation (SNSF - PCEFP2\_194658); United Kingdom: Leverhulme Trust (Leverhulme Trust RPG-2020-004), Royal Society (NIF-R1-231091); United States of America: U.S. Department of Energy (ECA DE-AC02-76SF00515), Neubauer Family Foundation.

%
%


%
%
%
\printbibliography
\clearpage
 
\begin{flushleft}
\hypersetup{urlcolor=black}
{\Large The ATLAS Collaboration}

\bigskip

\AtlasOrcid[0000-0002-6665-4934]{G.~Aad}$^\textrm{\scriptsize 104}$,
\AtlasOrcid[0000-0001-7616-1554]{E.~Aakvaag}$^\textrm{\scriptsize 17}$,
\AtlasOrcid[0000-0002-5888-2734]{B.~Abbott}$^\textrm{\scriptsize 123}$,
\AtlasOrcid[0000-0002-0287-5869]{S.~Abdelhameed}$^\textrm{\scriptsize 119a}$,
\AtlasOrcid[0000-0002-1002-1652]{K.~Abeling}$^\textrm{\scriptsize 56}$,
\AtlasOrcid[0000-0001-5763-2760]{N.J.~Abicht}$^\textrm{\scriptsize 50}$,
\AtlasOrcid[0000-0002-8496-9294]{S.H.~Abidi}$^\textrm{\scriptsize 30}$,
\AtlasOrcid[0009-0003-6578-220X]{M.~Aboelela}$^\textrm{\scriptsize 45}$,
\AtlasOrcid[0000-0002-9987-2292]{A.~Aboulhorma}$^\textrm{\scriptsize 36e}$,
\AtlasOrcid[0000-0001-5329-6640]{H.~Abramowicz}$^\textrm{\scriptsize 155}$
% ...
\bigskip
\\

% ...
$^{17}$Department for Physics and Technology, University of Bergen, Bergen; Norway.\\
% ...
$^{30}$Physics Department, Brookhaven National Laboratory, Upton NY; United States of America.\\
% ...
$^{36}$$^{(a)}$Facult\'e des Sciences Ain Chock, Universit\'e Hassan II de Casablanca;$^{(b)}$Facult\'{e} des Sciences, Universit\'{e} Ibn-Tofail, K\'{e}nitra;$^{(c)}$Facult\'e des Sciences Semlalia, Universit\'e Cadi Ayyad, LPHEA-Marrakech;$^{(d)}$LPMR, Facult\'e des Sciences, Universit\'e Mohamed Premier, Oujda;$^{(e)}$Facult\'e des sciences, Universit\'e Mohammed V, Rabat;$^{(f)}$Institute of Applied Physics, Mohammed VI Polytechnic University, Ben Guerir; Morocco.\\
% ...
$^{45}$Physics Department, Southern Methodist University, Dallas TX; United States of America.\\
% ...
$^{50}$Fakult\"{a}t Physik , Technische Universit{\"a}t Dortmund, Dortmund; Germany.\\
% ...
$^{56}$II. Physikalisches Institut, Georg-August-Universit\"{a}t G\"ottingen, G\"ottingen; Germany.\\
% ...
$^{104}$CPPM, Aix-Marseille Universit\'e, CNRS/IN2P3, Marseille; France.\\
% ...
$^{119}$$^{(a)}$New York University Abu Dhabi, Abu Dhabi;$^{(b)}$United Arab Emirates University, Al Ain; United Arab Emirates.\\
% ...
$^{123}$Homer L. Dodge Department of Physics and Astronomy, University of Oklahoma, Norman OK; United States of America.\\
% ...
$^{155}$Raymond and Beverly Sackler School of Physics and Astronomy, Tel Aviv University, Tel Aviv; Israel.\\
% ...
$^{a}$ Also Affiliated with an institute covered by a cooperation agreement with CERN.\\
$^{b}$ Also at An-Najah National University, Nablus; Palestine.\\
$^{c}$ Also at Borough of Manhattan Community College, City University of New York, New York NY; United States of America.\\
$^{d}$ Also at Center for High Energy Physics, Peking University; China.\\
$^{e}$ Also at Center for Interdisciplinary Research and Innovation (CIRI-AUTH), Thessaloniki; Greece.\\
$^{f}$ Also at Centro Studi e Ricerche Enrico Fermi; Italy.\\
% ...

\end{flushleft}


\clearpage

\end{document}